\documentclass[10pt,a4paper]{article}

\usepackage[margin=1.5in]{geometry}

\usepackage{graphicx}
\usepackage[linesnumbered,tworuled,vlined]{algorithm2e}
\usepackage{amsmath,theorem}
\usepackage{amssymb,dsfont}
\usepackage{bm,color}
\allowdisplaybreaks

\DeclareMathAlphabet{\mathonebb}{U}{bbold}{m}{n}

\newcommand{\R}{{\mathbb R}}

\newcommand{\D}[2]{ \ensuremath{ \frac{\mathrm{d} #1 }{\mathrm{d} #2 } }}
\newcommand{\DP}[2]{ \ensuremath{ \frac{\partial #1 }{\partial #2 } } }

\newcommand{\proba}[1]{\ensuremath{{\Bbb P}\!\left[#1 \right]} }
\newcommand{\esp}[1]{\ensuremath{{\Bbb E}\!\left[#1 \right]} }

\newtheorem{theorem}{Theorem}[section]
\newtheorem{lemma}[theorem]{Lemma}

\makeatletter
\newcommand*{\inlineequation}[2][]{%
  \begingroup
    \refstepcounter{equation}%
    \ifx\\#1\\%
    \else
      \label{#1}%
    \fi
    \relpenalty=10000 %
    \binoppenalty=10000 %
    \ensuremath{%
      #2%
    }%
    ~\@eqnnum
  \endgroup
}
\makeatother

\begin{document}

\abovedisplayskip=3pt
\belowdisplayskip=3 pt
\abovedisplayshortskip=0pt
\belowdisplayshortskip=0pt

\title{Effective computational methods for hybrid stochastic gene networks}

\author{Guilherme C.P. Innocentini$^1$, Fernando Antoneli$^2$, Arran Hodgkinson$^3$, \\
 Ovidiu Radulescu$^3$ \\
\small  $^1$ Federal University of ABC, Santo Andr\'e, Brazil, \\
\small  $^2$ Escola Paulista de Medicina, Universidade Federal de S\~ao Paulo, S\~ao Paulo, Brazil,\\
\small  $^3$ DIMNP UMR CNRS 5235, University of Montpellier, Montpellier, France.}

\maketitle

\sloppy

\begin{abstract}
At the scale of the individual cell, protein production is a stochastic process with multiple time scales, combining quick and slow random steps with discontinuous and smooth variation. Hybrid stochastic processes, in particular piecewise-deterministic Markov processes (PDMP), are well adapted for describing such situations. PDMPs approximate the jump Markov processes traditionally used as models for stochastic chemical reaction networks. Although hybrid modelling is now well established in biology, these models remain computationally challenging. We propose several improved methods for computing time dependent multivariate probability distributions (MPD) of PDMP models of gene networks. In these models, the promoter dynamics is described by a finite state, continuous time Markov process, whereas the mRNA and protein levels follow ordinary differential equations (ODEs). The Monte-Carlo method
combines direct simulation of the PDMP with analytic solutions of the ODEs.
The push-forward method numerically computes the probability measure advected by the deterministic ODE flow, through the use of analytic expressions of the corresponding semigroup. Compared to earlier versions of this method, the probability of the promoter states sequence is computed beyond the na\"ive mean field theory and adapted for non-linear regulation functions.

{\em Availability.} The algorithms described in this paper were implemented in MATLAB. The code is available on demand.
\end{abstract}

\section{Introduction}
In PDMP models of gene networks, each gene promoter is described as a finite state Markov process \cite{crudu2009hybrid,innocentini2018time,lin2018efficient,kurasov2018stochastic}.
The promoter triggers synthesis of gene products (mRNAs and proteins) with intensities depending on its state.
The promoter can exhibit two state (ON-OFF) dynamics, but also dynamics with more than two states and arbitrarily complex
transitions\cite{innocentini2013multimodality,thomas2014phenotypic}. The transition rates between the states of the promoter depend on the expression levels
of proteins expressed by the same or by other promoters.
In PDMP models, the gene products are considered in sufficiently large copy numbers
and are represented as continuous variables following ordinary differential equations (ODEs).
The sources of noise in these models are thus the discrete transitions between the promoter states.

In single cell experimental settings the quantities
of mRNA \cite{thattai2004stochastic,raj2006stochastic,cai2006stochastic,tantale2016single}
 and proteins \cite{elowitz2002stochastic,ferguson2012} can be determined
for each cell.
By double or multiple- fluorophore fluorescence techniques products from
several genes can be quantified simultaneously and one can have access to multivariate
probability distributions (MPD) of mRNA or proteins.
The stochastic dynamics of promoters and gene networks can have important
 consequences for
 fundamental biology \cite{eldar2010functional} but also for HIV  \cite{razooky2015hardwired}
and cancer research \cite{gupta2011stochastic}.
For this reason we aim to develop
effective methods for computing time-dependent MPDs for PDMP models.
Our main objective is the reduction of computation time which is prerequisite for parameter scans and machine learning applications \cite{herbach2017inferring}.

PDMPs already represent a gain with respect to the chemical Markov equation from
which they are derived by various limit theorems \cite{crudu2011convergence}. A gene network PDMP model
can be simulated by numerical integration of ODEs coupled with a driven
inhomogeneous Poisson process for the successive transitions
of the promoters \cite{zeiser2008simulation,crudu2009hybrid,riedler2013almost,lin2018efficient}. The simulation becomes particularly
effective when analytic solutions of the ODEs are available \cite{innocentini2018time}.

However, very little has been done to further
improve the computational power by optimising simulation and analysis of PDMP models.

Numerical integration of the PDE satisfied by MPD is an interesting option combining precision
and speed for small models. Finite difference methods, however, are of limited use in this context as they can not cope with
many RNA and protein variables (extant examples are restricted to the dimension 2, corresponding
to a single promoter, with or without self-regulation
see \cite{innocentini2018time,kurasov2018stochastic}).

Another interesting method for computing time dependent MPDs is the push-forward method.
For gene networks, this method has been first introduced in \cite{innocentini2016protein}
and further adapted for continuous mRNA variables in \cite{innocentini2018time}.
It is based on the idea to compute the MPD as the push-forward measure of the semigroup
defined by the ODEs. This method is approximate, as one has to consider that the
discrete PDMP variables are piecewise constant on a deterministic time partition. Furthermore,
the transition rates between promoter states were computed in a mean field approximation.
In this paper we replace the mean field approximation by the next order approximation
taking into account the moments of the protein distribution.

\section{Methods}

\subsection{PDMP models of gene networks}

The state of a PDMP gene network model takes values in $E = \R^{2N} \times \{0,\ldots,s_{max}\}^{N}-1$,
where $N$ is the number of genes and $s_ {max}$ is a positive integer representing the
maximum number of states of a gene promoter. It is a process $\vec{\zeta}_t = (\vec{x}_t,\vec{y}_t,\vec{s}_t)$, determined by three characteristics:
\begin{description}
\item[1)]
For all $\vec{s} \in \{0,\ldots,s_{max}\}^{N}$ a vector field
 $\vec{F}_{\vec{s}} : \R^{2N} \to \R^{2N} $ determining
a unique global flow $\vec{\Phi}_{\vec{s}}(t,\vec{x},\vec{y})$ in $\R^{2N}$, the space of all protein ($\vec{x} \in \R^{N}$)
 and mRNA ($\vec{y} \in \R^{N}$) values such that, for $t>0$,
\begin{equation}
\D{\vec{\Phi}_{\vec{s}}(t,\vec{x},\vec{y})}{t} =  \vec{F}_{\vec{s}}(\vec{\Phi}_{\vec{s}}(t,\vec{x},\vec{y})), \;
\vec{\Phi}_{\vec{s}}(0,\vec{x},\vec{y})=(\vec{x},\vec{y}). \label{flow}
\end{equation}
On coordinates, this reads
\begin{eqnarray}
\D{\Phi_i^x}{t}& = &b_i \Phi_i^y - a_i \Phi_i^x, \notag \\
\D{\Phi_i^y}{t} &= & k_i(s_i) - \rho_i \Phi_i^y, \; 1 \leq i \leq N, \label{standard1}
\end{eqnarray}
where $b_i$, $k_i$, $a_i$, $\rho_i$ are translation efficiencies, transcription rates, protein
degradation coefficients and mRNA degradation coefficients of the $i^\text{th}$ gene, respectively.
Note that transcription rates depend on the relevant promoter states.

The flow $\vec{\Phi}_{\vec{s}}(t,\vec{x},\vec{y})$ represents a one parameter semigroup
fulfilling the properties
\begin{description}
\item[(i)]
$\vec{\Phi}_{\vec{s}}(0,\vec{x}_0,\vec{y}_0) = (\vec{x}_0,\vec{y}_0)$,
\item[(ii)]
$\vec{\Phi}_{\vec{s}}(t+t',\vec{x}_0,\vec{y}_0) = \vec{\Phi}_{\vec{s}}(t',
\vec{\Phi}_{\vec{s}}^x(t,\vec{x}_0,\vec{y}_0),
\vec{\Phi}_{\vec{s}}^y(t,\vec{x}_0,\vec{y}_0))$.
\end{description}
\item[2)]
A transition rate matrix for the promoter states $\vec{H} : \R^{2N} \to  M_{N\times N} (\R)$, such
that $H_{\vec{s},\vec{r}}(\vec{x},\vec{y}) \geq 0$  and
$H_{\vec{s},\vec{s}}(\vec{x},\vec{y}) =  - \sum_{\vec{r} \neq \vec{s}} H_{\vec{r},\vec{s}}(\vec{x},\vec{y})$ for all $\vec{s},\vec{r}\in \{0,\ldots,s_{max}\}^{N}, \vec{s}\neq \vec{r}$ and for all $(\vec{x},\vec{y})\in R^{2N}$.
\item[3)]
A jump rate $\lambda : E \to \R^+$. The jump rate can be obtained from the transition rate matrix
\begin{equation}\label{transition}
\lambda(\vec{x},\vec{y},\vec{s}) = \sum_{\vec{r}\neq \vec{s}} H_{\vec{r},\vec{s}}(\vec{x},\vec{y}) = - H_{\vec{s},\vec{s}}(\vec{x},\vec{y}).
\end{equation}
\end{description}
From these characteristics, right-continuous sample paths
$\{(\vec{x}_t,\vec{y}_t) : t >0 \}$ starting at $\vec{\zeta}_0=(\vec{x}_0,\vec{y}_0,\vec{s}_0) \in E$
can be constructed as follows. Define
\begin{equation}
\vec{x}_t (\omega) := \vec{\Phi}_{\vec{s}_0}(t,\vec{x}_0,\vec{y}_0)
\text{ for } 0 \leq t \leq T_1(\omega),
\end{equation}
where $T_1(\omega)$ is a realisation of the first jump time of $\vec{s}$, with the distribution
\begin{equation}
F(t) = \proba{T_1 > t} = \exp (-\int_{0}^{t} \lambda(\vec{\Phi}_{\vec{s}_0}(u,\vec{x}_0,\vec{y}_0)) d u), \; t>0, \label{next}
\end{equation}
and $\omega$ is the element of the probability space for which the particular realisation of the process is given.
The pre-jump state is $\vec{\zeta}_{T_1^-(\omega)} (\omega) = (\vec{\Phi}_{\vec{s}_0}(T_1(\omega),\vec{x}_0,\vec{y}_0), \vec{s}_0 )$ and the post-jump state
is  $\vec{\zeta}_{T_1(\omega)} (\omega) = (\vec{\Phi}_{\vec{s}_0}(T_1(\omega),\vec{x}_0,\vec{y}_0), \vec{s} )$, where $\vec{s}$ has the distribution
\begin{equation}
\proba{\vec{s}=\vec{r}} = \frac{H_{\vec{r},\vec{s_0}} (\vec{\Phi}_{\vec{s}_0}(T_1(\omega),\vec{x}_0,\vec{y}_0),\vec{s_0}) }{ \lambda(\vec{\Phi}_{\vec{s}_0}(T_1(\omega),\vec{x}_0,\vec{y}_0),\vec{s_0})}, \text{ for all }
\vec{r} \neq \vec{s}_0.
\end{equation}
We then restart the process  $\vec{\zeta}_{T_1(\omega)}$ and recursively apply the same procedure
at jump times $T_2(\omega)$, etc..

Note that between each two consecutive jumps $(\vec{x}_t,\vec{y}_t)$ follow deterministic ODE dynamics
defined by the vector field $\vec{F}_{\vec{s}}$. At the jumps, the protein and mRNA values
  $(\vec{x}_t,\vec{y}_t)$  are continuous.

The calculation of the flow between two jumps and of the jump time can be gathered in
the same set of differential equations
\begin{eqnarray}
\D{\Phi_i^x}{t}& = &b_i \Phi_i^y - a_i \Phi_i^x, \notag \\
\D{\Phi_i^y}{t} &= & k_i(s_i) - \rho_i \Phi_i^y, \; 1 \leq i \leq N, \notag \\
\D{\log F}{t} &=& - \lambda(\vec{x},\vec{y},\vec{s_0}),
\label{standard2}
\end{eqnarray}
that has to be integrated with the stopping condition $F(T_1) = U$, where $U$ is
a random variable, uniformly distributed on $[0,1]$.

We define multivariate probability density functions $p_{\vec{s}}(t,\vec{x},\vec{y})$.
These functions satisfy the Liouville-master equation which is a system
of partial differential equations:
\begin{equation}
\DP{p_{\vec{s}}(t,\vec{x},\vec{y})}{t} =
- \nabla_{\vec{x},\vec{y}} . (\vec{F}_{\vec{s}}(\vec{x},\vec{y}) p_{\vec{s}}(t,\vec{x},\vec{y}) )
+  \sum_{\vec{r}} H_{\vec{s},\vec{r}}(\vec{s},\vec{x},\vec{y}) p_{\vec{r}}(t,\vec{x},\vec{y}). \label{master-liouville}
\end{equation}

\subsection{ON/OFF gene networks}
In this paper, for the purpose of illustration only, all the examples are constituted by ON/OFF gene networks.

For an ON/OFF gene each component $s_i$ has two possible values $0$ for OFF and $1$
for ON.

As a first example that we denote as model $M1$,
let us consider a two genes network; the expression of the
first gene being constitutive and the expression of the second gene being
activated by the first. We consider that the transcription activation
rate of the second gene is proportional to the concentration of the first
protein $f_2 x_1$. All the other rates are constant $f_1$, $h_1$, $h_2$,
representing the transcription activation rate of the first gene, and the
transcription inactivation rates of gene one and gene two, respectively.
For simplicity, we consider that the two genes have identical
protein and mRNA parameters $b_1 = b_2 = b$, $a_1 = a_2 = a$,
$\rho_1 = \rho_2 = \rho$.
 We further consider that
$k_i(s_i) = k_0$ if the gene $i$ is OFF and $k_i(s_i) = k_1$ if the gene $i$ is ON.

The gene network has four discrete states, in order $(0,0)$, $(1,0)$, $(0,1)$, and $(1,1)$.
Then, the transition rate matrix for the model $M_1$ is
\begin{equation}
\begin{bmatrix}
-(f_1 + f_2x_1) & h_1 & h_2 & 0 \\
f_1 & -(h_1+f_2x_1) & 0 & h_2  \\
f_2x_1 & 0 & -(f_1 + h_2) & h_1 \\
0 & f_2x_1  & f_1 & -(h_1+h_2)
\end{bmatrix}.
\end{equation}
The Liouville-master equation for the model $M1$ reads
\begin{eqnarray}
\DP{p_1}{t}  &= &
-\DP{[(b y_1 -  a x_1) p_1]}{x_1} - \DP{ [(k_0 - \rho y_1) p_1]}{y_1}
 -\DP{[(b y_2 -  a x_2) p_1]}{x_2} - \DP{ [(k_0 - \rho y_2) p_1]}{y_2} +\notag \\
 &+&   h_2p_3+h_1p_2 -  (f_1+f_2x_1)  p_1,\notag \\
\DP{p_2}{t}  &= &
-\DP{[(b y_1 -  a x_1) p_2]}{x_1} - \DP{ [(k_1 - \rho y_1) p_2]}{y_1}
 -\DP{[(b y_2 -  a x_2) p_2]}{x_2} - \DP{ [(k_0 - \rho y_2) p_2]}{y_2} + \notag \\
 &+&   f_1 p_1+ h_2 p_4 -  (h_1+f_2x_1)  p_2, \notag \\
\DP{p_3}{t}  &= &
-\DP{[(b y_1 -  a x_1) p_3]}{x_1} - \DP{ [(k_0 - \rho y_1) p_3]}{y_1}
 -\DP{[(b y_2 -  a x_2) p_3]}{x_2} - \DP{ [(k_1 - \rho y_2) p_3]}{y_2} + \notag \\
 &+&   h_1 p_4+ f_2x_1 p_1 -  (h_2+f_1)  p_3, \notag \\
\DP{p_4}{t}  &= &
-\DP{[(b y_1 -  a x_1) p_4]}{x_1} - \DP{ [(k_1 - \rho y_1) p_4]}{y_1}
 -\DP{[(b y_2 -  a x_2) p_4]}{x_2} - \DP{ [(k_1 - \rho y_2) p_4]}{y_2} + \notag \\
 &+&   f_1 p_3+ f_2x_1 p_2 -  (h_1+h_2)  p_4.
  \label{liouville-master-twogene}
\end{eqnarray}
The model $M_2$ differs from the model $M_1$ by the form of the activation function. Instead of a linear
transcription rate $f_2 x_1$ we use a Michaelis-Menten model $f_2 x_1/(K_1 + x_1)$. This model is
more realistic as it takes into account that the protein $x_1$ has to attach to
specific promoter sites which become saturated when the concentration
of this protein is high.

The transition rate matrix for the model $M_2$ is
\begin{equation}
\begin{bmatrix}
-(f_1 + f_2x_1/(K_1+x_1)) & h_1 & h_2 & 0 \\
f_1 & -(h_1+f_2x_1/(K_1+x_1)) & 0 & h_2  \\
f_2x_1/(K_1+x_1) & 0 & -(f_1 + h_2) & h_1 \\
0 & f_2x_1/(K_1+x_1)  & f_1 & -(h_1+h_2)
\end{bmatrix}.
\end{equation}
The Liouville-master equation for the model $M2$ reads
\begin{eqnarray}
\DP{p_1}{t}  &= &
-\DP{[(b y_1 -  a x_1) p_1]}{x_1} - \DP{ [(k_0 - \rho y_1) p_1]}{y_1}
 -\DP{[(b y_2 -  a x_2) p_1]}{x_2} - \DP{ [(k_0 - \rho y_2) p_1]}{y_2} +\notag \\
 &+&   h_2p_3+h_1p_2 -  (f_1+f_2x_1/(K_1+x_1))  p_1,\notag \\
\DP{p_2}{t}  &= &
-\DP{[(b y_1 -  a x_1) p_2]}{x_1} - \DP{ [(k_1 - \rho y_1) p_2]}{y_1}
 -\DP{[(b y_2 -  a x_2) p_2]}{x_2} - \DP{ [(k_0 - \rho y_2) p_2]}{y_2} + \notag \\
 &+&   f_1 p_1+ h_2 p_4 -  (h_1+f_2x_1/(K_1+x_1))  p_2, \notag \\
\DP{p_3}{t}  &= &
-\DP{[(b y_1 -  a x_1) p_3]}{x_1} - \DP{ [(k_0 - \rho y_1) p_3]}{y_1}
 -\DP{[(b y_2 -  a x_2) p_3]}{x_2} - \DP{ [(k_1 - \rho y_2) p_3]}{y_2} + \notag \\
 &+&   h_1 p_4+ f_2x_1/(K_1+x_1) p_1 -  (h_2+f_1)  p_3, \notag \\
\DP{p_4}{t}  &= &
-\DP{[(b y_1 -  a x_1) p_4]}{x_1} - \DP{ [(k_1 - \rho y_1) p_4]}{y_1}
 -\DP{[(b y_2 -  a x_2) p_4]}{x_2} - \DP{ [(k_1 - \rho y_2) p_4]}{y_2} + \notag \\
 &+&   f_1 p_3+ f_2x_1/(K_1+x_1) p_2 -  (h_1+h_2)  p_4.
  \label{liouville-master-twogene}
\end{eqnarray}

\subsection{Monte-Carlo method}
The Monte-Carlo method utilizes the direct simulation of the PDMP based on Eq.\ref{standard2}.
A larger number $M$ of sample paths is generated and the values of $(\vec{x}_t,\vec{y}_t)$ are
stored at selected times. Multivariate probability distributions are then estimated from this
data.

The direct simulation of PDMPs
needs the solutions of \eqref{standard2} which can be obtained by numerical
integration. This is not always computationally
easy.
Problems may arise for fast switching
promoters when the ODEs have to be integrated many times on small intervals between
successive jumps.
Alternatively, the numerical integration of the ODEs can be replaced by analytic
solutions or quadratures. Analytic expressions are always available for the gene
network flow \eqref{standard1} and read
\begin{eqnarray}
\Phi_i^x(t,x_0,y_0) &=&
x_0 \exp ( - a_i t) + b_i \left[\left(y_0 - \frac{k_i(s_i)}{\rho_i}\right)\frac{\exp (-\rho_i t)-1}{a_i-\rho_i}+ \frac{k_i(s_i)}{\rho_i}\frac{1-\exp(-a_i t)}{a_i} \right], \notag \\
\Phi_i^y(t,x_0,y_0) &=& (y_0 - k_i/\rho_i)\exp ( - \rho_i t) + k_i/\rho_i.
\end{eqnarray}
Let us consider the following general expression of the jump intensity function
$$\lambda(\vec{x},\vec{y},\vec{s}) = c_0(\vec{s}) +
\sum_{i}^N c_i (\vec{s}) x_i + \sum_{i}^N d_i (\vec{s}) f_i(x_i),$$
where $f_i$ are non-linear functions, for instance Michaelis-Menten
$f_i(x_i) = x_i / (K_i + x_i)$ or Hill functions $f_i(x_i) = x_i^{n_i} / (K_i^{n_i} + x_i^{n_i})$.
If  $d_i=0$ for all $1\leq i \leq N$, the cumulative distribution function of the waiting time $T_1$
can be solved analytically \cite{innocentini2018time}, otherwise it can be obtained by quadratures. For example,
for the model $M_2f$ one has
$$\lambda(\vec{x},\vec{y},\vec{s}) = \left(f_1 + f_2 \frac{x_1}{K_1 + x_1}\right)\delta_{s,1} +
\left(h_1 + f_2 \frac{x_1}{K_1 + x_1}\right) \delta_{s,2} + (h_2 + f_1) \delta_{s,3} +
(h_2 + h_1) \delta_{s,4},
$$
where $\delta_{i,j}$ is Kronecker's delta.
In this case the
 waiting time $T_1$ is obtained as the unique solution of the equation
\begin{equation}
\begin{split}
&-\log (U) = \bigg[(f_1 + f_2)T_1 +f_2 \int_0^{T_1} \frac{1}{K_1 + \Phi_1^x(t',x_0,y_0)}\, dt' \bigg] \delta_{s_0,1} +
\bigg[(h_1 + f_2)T_1 + \\ & f_2\int_0^{T_1} \frac{1}{K_1 + \Phi_1^x(t',x_0,y_0)} \, dt' \bigg] \delta_{s_0,2}
 +  (h_2 + f_1)T_1 \delta_{s_0,3} + (h_2 + h_1)T_1 \delta_{s_0,4},
\end{split}\label{T1}
\end{equation}
where $U$ is a random variable, uniformly distributed on $[0,1]$.
In our implementation of the algorithm we solve \eqref{T1} numerically, using
the bisection method.

%

\subsection{Push-forward method}

This method allows one to compute the MPD of proteins and mRNAs at a time $\tau$ given the MPD of proteins and mRNAs at time $0$.

In order to achieve this we use a deterministic partition $\tau_0=0 < \tau_1 < \ldots < \tau_M=\tau$ of the interval $[0,\tau]$
such that $\Delta_M = \max_{j\in[1,M]} (\tau_{j} - \tau_{j-1})$ is small.
The main approximation of this method is to consider that $s(t)$ is piecewise constant on this partition, more
precisely that $s(t) = s_j := s(\tau_j), \, \text{for } t \in [\tau_j,\tau_{j+1}), \, 0 \leq j \leq M-1$. This approximation is justified by Theorem~\ref{theorem1} in Section~\ref{sec:convergence}.

For each path realisation $S_M:=(s_0,s_1,\ldots,s_{M-1})\in \{0,1,\ldots,2^N-1\}^M$ of the promoter states, we can compute (see Appendix 2)
the protein and mRNA levels $x(\tau),y(\tau)$ of all genes $i \in [1,N]$:
\begin{eqnarray}
y_i(\tau) &=& y_i(0)e^{-\rho\tau} + \frac{k_0}{\rho} (1-e^{- \rho \tau}) + \frac{k_1-k_0}{\rho}\sum_{j=1}^{M-1}e^{-\rho \tau} (e^{-\rho \tau_{j+1}}-e^{-\rho \tau_j})s_j^i \label{eqyt} \\
x_i(\tau) &=& x_i(0) e^{-a \tau} + \frac{b y_i(0)}{a-\rho} (e^{-\rho \tau} - e^{-a \tau}) + \frac{bk_0}{\rho}\left(\frac{1-e^{-a\tau}}{a}+\frac{e^{-a\tau}-e^{-\rho \tau}}{a-\rho}\right) +\notag \\
&+& \frac{b (k_1-k_0)}{\rho}e^{-a\tau}\sum_{j=1}^M s_{j-1}^i w_{j}, \, i \in [1,N] \label{eqxt}
\end{eqnarray}
where
$w_{j}=\frac{e^{(a-\rho)\tau}-e^{(a-\rho)\tau_j}}{a-\rho}(e^{\rho \tau_j}-e^{\rho \tau_{j-1}}) - \frac{e^{(a-\rho)\tau_j}-e^{(a-\rho)\tau_{j-1}}}{a-\rho}e^{\rho\tau_{j-1}}
+\frac{e^{a\tau_j}-e^{a\tau_{j-1}}}{a}$ and $s_j^i: = 0$ if promoter $i$ is OFF for
$t \in [\tau_j,\tau_{j+1})$ and $s_j^i: = 1$ if promoter $i$ is ON for
$t \in [\tau_j,\tau_{j+1})$.

In order to compute the MDP at time $\tau$ one has to sum the contributions of all
solutions
\eqref{eqyt},\eqref{eqxt}, obtained for the
 $2^{NM}$ realisations of promoter state paths
with weights given by the probabilities of the paths.

Eqs.\ref{eqyt},\ref{eqxt} can straightforwardly be adapted to
compute $x(t),y(t)$ for all $t \in [0,\tau]$. To this aim, $\tau$ should be
replaced by $t$ and $M$ should be replaced by $M_t$ defined by the relation
$t \in [\tau_{M_t},\tau_{{M_t}+1}]$.

Suppose that we want to estimate the MDP of all mRNAs and proteins of the
gene network, using a multivariate histogram with bin centers
$(x_0^{l_i},y_0^{m_i}), \, 1 \leq i \leq N, 1\leq l_i \leq n_x,  1\leq m_i \leq n_y$ where $n_x$, $n_y$ are the numbers of bins in the protein and mRNA directions for each gene, respectively. Typically
$x_0^{l_i} = b/(a\rho) (k0 + (k1-k0)(l_i-1/2) ),1 \leq i \leq N, 1 \leq l_i \leq n_x $,
$y_0^{m_i} =   1/\rho (k0 + (k1-k0)(m_i-1/2) ),1 \leq i \leq N, 1 \leq m_i \leq n_y$.
The initial MDP at time $t=0$ is given by the bin probabilities
$p_0^{l_i,m_i},1 \leq i \leq N, 1\leq l_i \leq  n_x, 1 \leq m_i \leq n_y$.
Let $(x^{l_i,m_i},y^{l_i,m_i})$ be the solutions \label{xt},\label{yt} with
$x_i(0)=x_0^{l_i}$ and $y_i(0)=y_0^{m_i}$. The many-to-one
application $(l_i',m_i') = \psi (l_i,m_i)$ provides the histogram bin $(l_i',m_i')$ in which falls the vector
$(x^{l_i,m_i},y^{l_i,m_i}))$. The push forward MDP at time $t=\tau$ is defined by the
bin probabilities $p^{l_i,m_i}, 1 \leq i \leq N, 1\leq l_i \leq  n_x, 1 \leq m_i \leq n_y$ that are computed as
\begin{equation}
p^{l_i,m_i} = \sum_{S_M} \sum_{\psi(l'_i,m'_i)=(l_i,m_i)} p_0^{l'_i,m'_i} \proba{S_M}. \label{distribution}
\end{equation}

In order to compute  $\proba{S_M}$ we can use the fact that, given $x(t)$, $s(t)$
is a finite state Markov process, therefore
\begin{equation}\label{SM}
\proba{S_M} = \Pi_{s_{N-1},s_{N-2}}(\tau_{N-2},\tau_{N-1}) \ldots \Pi_{s_1,s_0}(\tau_0,\tau_1) P_0^S(s_0),
\end{equation}
where $P_0^S: \{ 0 , 1,\ldots, 2^N-1\} \to [0,1]$ is the initial distribution of the promoter state,
\begin{equation}
\vec{\Pi}(\tau_{j},\tau_{j+1}) = \exp \left( \int_{\tau_{j}}^{\tau_{j+1}} \vec{H}(\vec{x}(t)) \, dt \right),
\end{equation}
and $\vec{x}(t)$ are given by \eqref{eqxt}.

The push-forward method can be applied recursively to compute the MDP for times $\tau, \, 2\tau, \ldots, n_t \tau$.
The complexity of the calculation scales as $n_t (n_x)^N (n_y)^N 2^{NM}$ which is exponential in the number of
genes $N$. The exponential complexity  comes from considering all the $2^{NM}$ possible paths $S_M$. However,
many of these paths have almost the same probability and impose very similar trajectories to the variables
$(\vec{x}(t),\vec{y}(t))$. In fact, a convenient approximation is to consider that different genes
are switching between ON and OFF states according to Markov processes with rates given by the
mean values of regulatory proteins (mean field approximation, \cite{innocentini2018time}).
This approximation consists in applying the push-forward procedure for
each gene separately, using averaged transition probabilities. Thus,
the $2^N$ states transition matrix $\vec{H}$ has to be replaced by $N$, $2\times 2$  state transition matrices for each gene.
This approximation reduces the complexity of the calculations
to  $n_t n_x n_y N 2^M$ which is linear in the number of genes.

In \cite{innocentini2018time} we have replaced the regulation term $f_2 x_1(t)$ occurring in the transition matrix
by its mean $f_2 \esp{x_1(t)}$. In this case both $\vec{H}$ and $\vec{\Pi}$ can be computed analytically, which
leads to a drastic reduction in the execution time. This approach is suitable for the model $M_1$, which contains only linear regulation terms.
For non-linear regulation terms, $\vec{\Pi}$ can not generally be computed analytically. Furthermore, the mean field approximation introduces biases.
For instance, in the case of the model $M_2$, the approximation
$f_2 x_1(t)/(K_1+x_1(t)) \approx f_2 \esp{x_1(t)}/(K_1+\esp{x_1(t)})$ is poor. A better approximation in this case is to replace  $f_2 x_1(t)/(K_1+x_1(t))$ by its mean and use
\begin{equation}
\esp{\frac{f_2 x_1(t)}{K_1+x_1(t)}} \approx  \frac{f_2 \esp{x_1(t)}}{(K_1+\esp{x_1(t)})} - \frac{f_2 }{(K_1+\esp{x_1(t)})^3} Var(x_1(t)),
\end{equation}
in order to correct the bias. Here $Var$ indicates the variance.

As in \cite{innocentini2018time} we can use analytic expressions for $\esp{x_1(t)}$, but also for $Var(x_1(t))$. These expressions can be found in Appendix~1. Although the elements of matrix $\vec{H}$ have analytic expressions, the
elements of the matrix $\vec{\Pi}$ contain integrals that must be computed numerically. For the model $M_2$, we have
\begin{equation}
\vec{\Pi}_1(\tau,\tau') =
\begin{bmatrix}
(1-p_{1,on}) + p_{1,on}e^{-\epsilon_1 (\tau'-\tau)}, ~\ & (1-p_{1,on}) - (1-p_{1,on})e^{-\epsilon_1 (\tau'-\tau)} \\[.4em]
p_{1,on} - p_{1,on}e^{-\epsilon_1 (\tau'-\tau)}, & p_{1,on} + (1-p_{1,on})e^{-\epsilon_1 (\tau'-\tau)}  \\
\end{bmatrix},
\end{equation}
for the transition rates of the first gene, where $p_{1,on} = f_1/(f_1+h_1)$, $\epsilon_1 = (f_1+h_1)/\rho$, and
\begin{equation}
\begin{aligned}&
\vec{\Pi}_2(\tau,\tau') =	\\& ~\
\begin{bmatrix}
e^{-\int_{\tau}^{\tau'}(h_2 + F_2(t))\, dt} + h_2 \int_{\tau}^{\tau'} e^{-\int_{t}^{\tau'}(h_2+F_2(t'))\, dt'}\,dt, &
h_2 \int_{\tau}^{\tau'} e^{-\int_{t}^{\tau'}(h_2+F_2(t'))\, dt'}\,dt\\
1- e^{-\int_{\tau}^{\tau'}(h_2 + F_2(t))\, dt} - h_2 \int_{\tau}^{\tau'} e^{-\int_{t}^{\tau'}(h_2+F_2(t'))\, dt'}\,dt, ~\ &
1 - h_2 \int_{\tau}^{\tau'} e^{-\int_{t}^{\tau'}(h_2+F_2(t'))\, dt'}\,dt
\end{bmatrix},
\end{aligned}\label{pi2}
\end{equation}
for the transitions of the second gene, where $F_2(t) = f_2 \esp{\frac{x_1(t)}{K_1+x_1(t)}}$.


\section{Results}

\subsection{Convergence of the push-forward method}\label{sec:convergence}
The probability distribution obtained with the push-forward method converges to the exact PDMP distribution in the limit $M \to \infty$. This is a consequence of the
following theorem
\begin{theorem}\label{theorem1}
Let $\Phi_{S_M}(t,\vec{x},\vec{y})$ be the flow defined by the formulas \eqref{eqxt},\eqref{eqyt}, such that $(\vec{x}(t),\vec{y}(t))= \Phi_{S_M}(t,\vec{x}(0),\vec{y}(0))$ for $t\in[0,\tau]$, and let $\mu_t^M : \mathcal{B}(\R^{2N})\to \R_+$ be the probability measure defined as $\mu_t^M (A) = \sum_{S_M} \proba{S_M} \mu_0(\Phi_{S_M}^{-1}(t,A))$, where $\mu_0 : \mathcal{B}(\R^{2N})\to \R_+$ is
the probability distribution of $(\vec{x},\vec{y})$ at $t=0$,
$\proba{S_M}$ are given by \eqref{SM},
 and $\mathcal{B}(\R^{2N})$ are the Borel sets on $\R^{2N}$.
Let $\mu_t$,  the exact distribution of $(\vec{x}(t),\vec{y}(t))$ for the
PDMP defined by \eqref{flow},\eqref{standard1},\eqref{transition},
with initial values $(\vec{x}_0,\vec{y}_0,\vec{s}_0)$ distributed according to $\mu_0 \times P_0^S$.
 Assume that $|\tau_{i}-\tau_{i-1}| < C/M$
for all $i\in [1,M]$, where $C$ is a positive constant.
Then, for all $t\in [0,\tau]$,
 $\mu_t^M$  converges in distribution to  $\mu_t$, when $M \to \infty$.
\end{theorem}
The proof of this theorem is given in the Appendix 3.

\subsection{Testing the push-forward method}
In order to test the push-forward method, we compared the resulting
probability distributions with the ones obtained by the Monte Carlo method
using the
direct simulation of the PDMP.
We considered the models $M_1$ and $M_2$ with the following
parameters: $\rho=1$, $p1=1/2$, $a=1/5$, $b=4$,
$k_0=4$, $k_1=40$ for the two genes.
The parameter $\epsilon$ took two values
$\epsilon=0.5$ for slow genes and $\epsilon=5.5$
for fast genes. We tested  the
slow-slow and the fast-fast combinations of parameters.

The initial distribution of the promoters states was $P_0^S( (0,0) )=1$ where
the state $(0,0)$ means that both promoters are OFF.
The initial probability measure $\mu_0$ was a delta Dirac
distribution centered at $x_1 = x_2=0$ and $y_1=y_2=0$. This is
obtained by always starting the direct simulation
of the PDMP from $x_1(0)=x_2(0)=0$,
$y_1(0)=y_2(0)=0$, and $s_1(0)= s_2(0) = 0$.
The simulations were performed between $t_0=0$ and $t_{max}=20$ for fast genes
and between $t_0=0$ and $t_{max}=90$ for slow genes. In order to estimate the
distributions we have used $MC=50000$ samples.

The push-forward method was implemented with
$M=10$ equal length sub-intervals of $[0,\tau]$.
The time step $\tau$ was chosen $\tau=2$ for fast genes and $\tau=15$ for slow genes. The procedure
was iterated $10$ times for fast genes (up to
$t_{max}=20$) and $6$ times for slow genes
(up to $t_{max}=90$).

\begin{figure}[h!]

\includegraphics[width=0.5\textwidth]{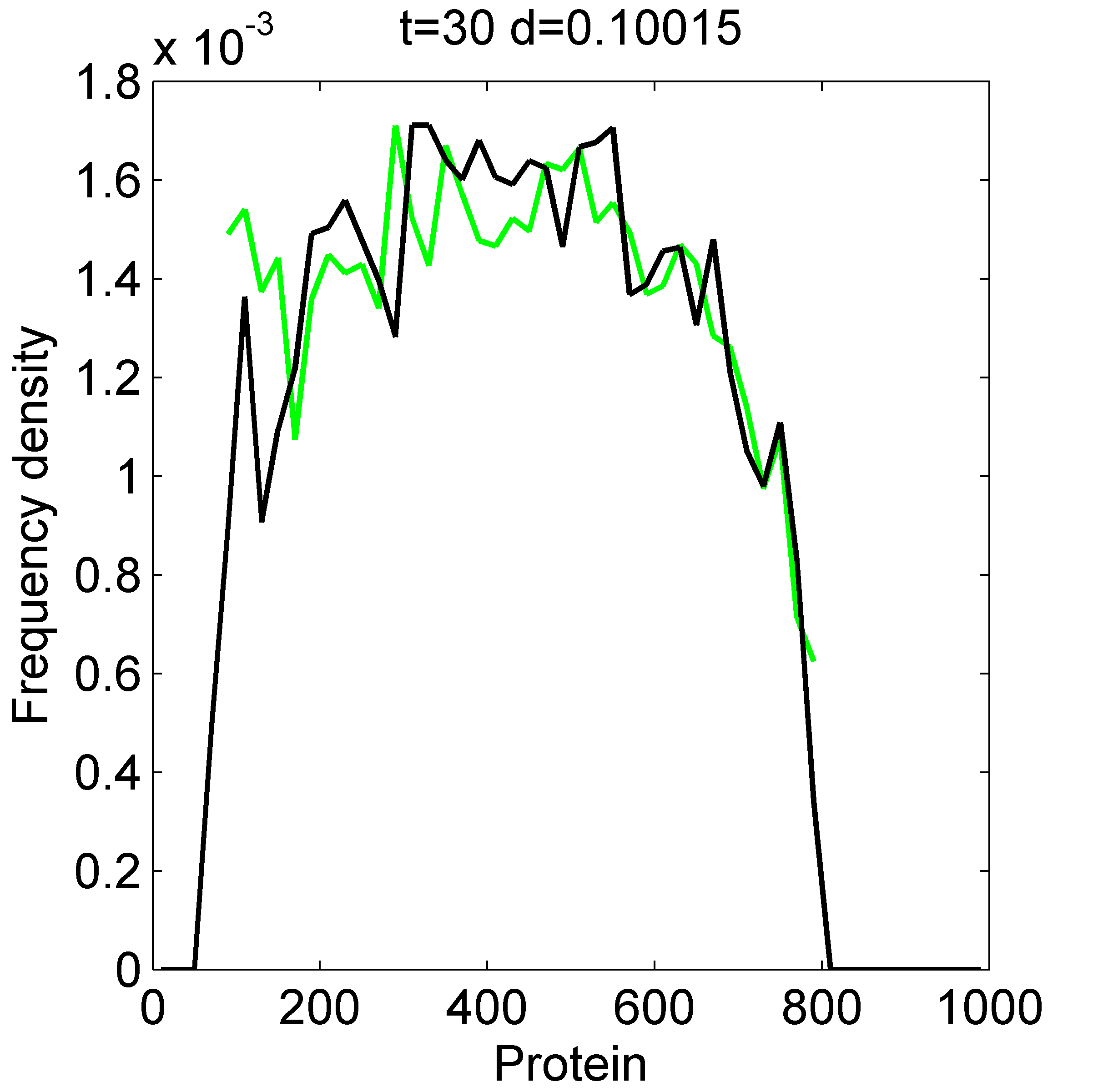}
\includegraphics[width=0.5\textwidth]{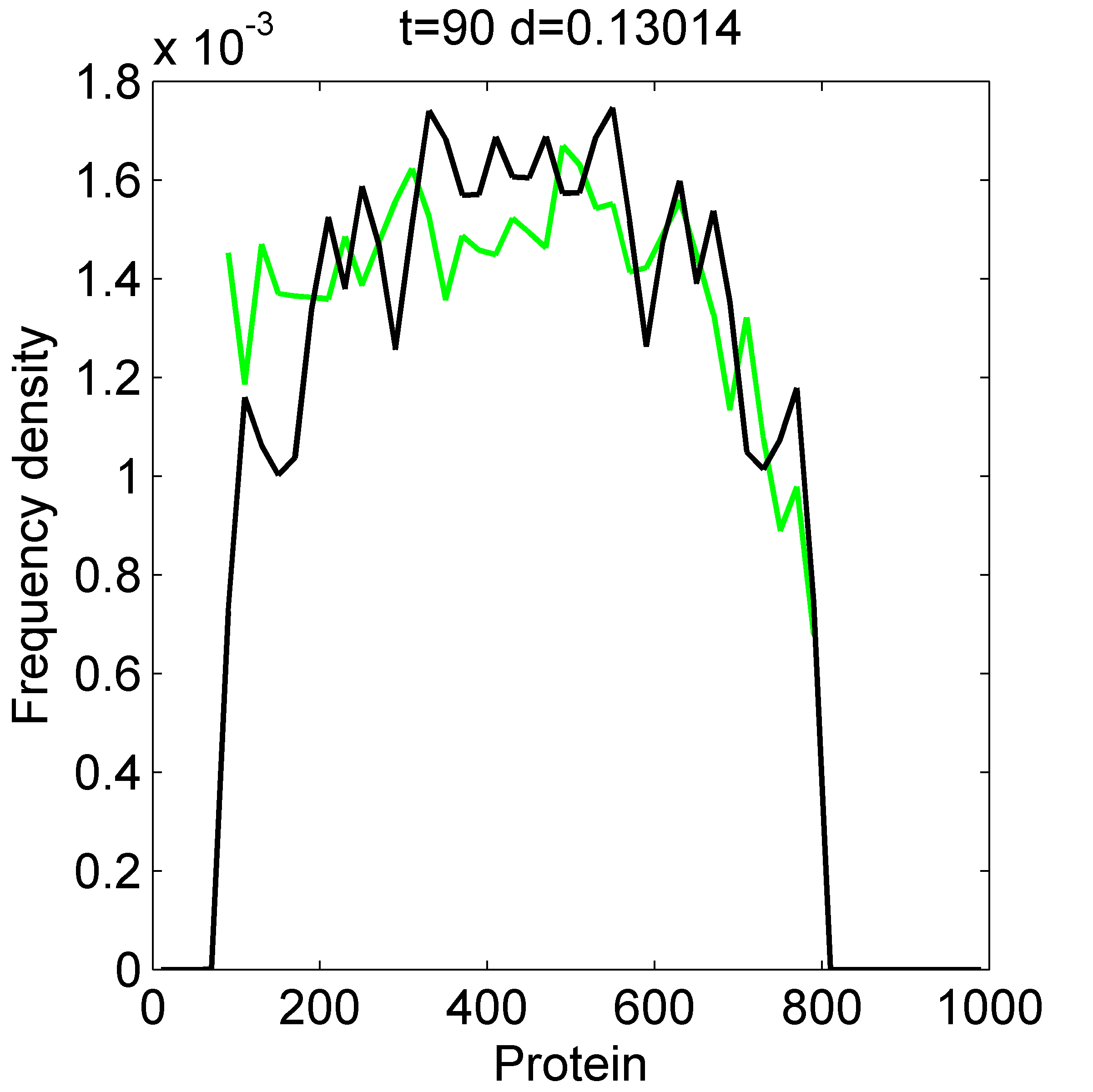}
\includegraphics[width=0.5\textwidth]{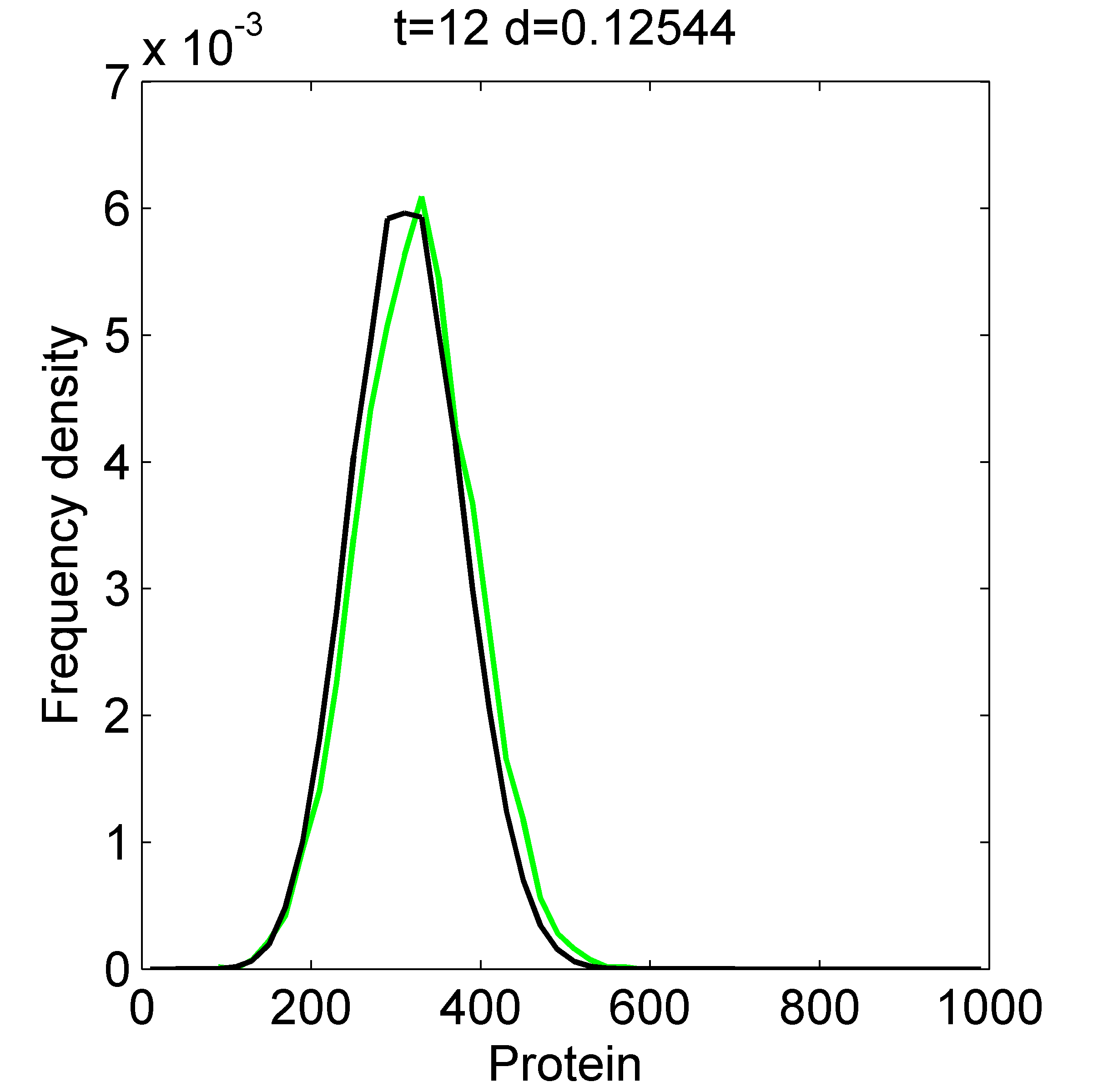}
\includegraphics[width=0.5\textwidth]{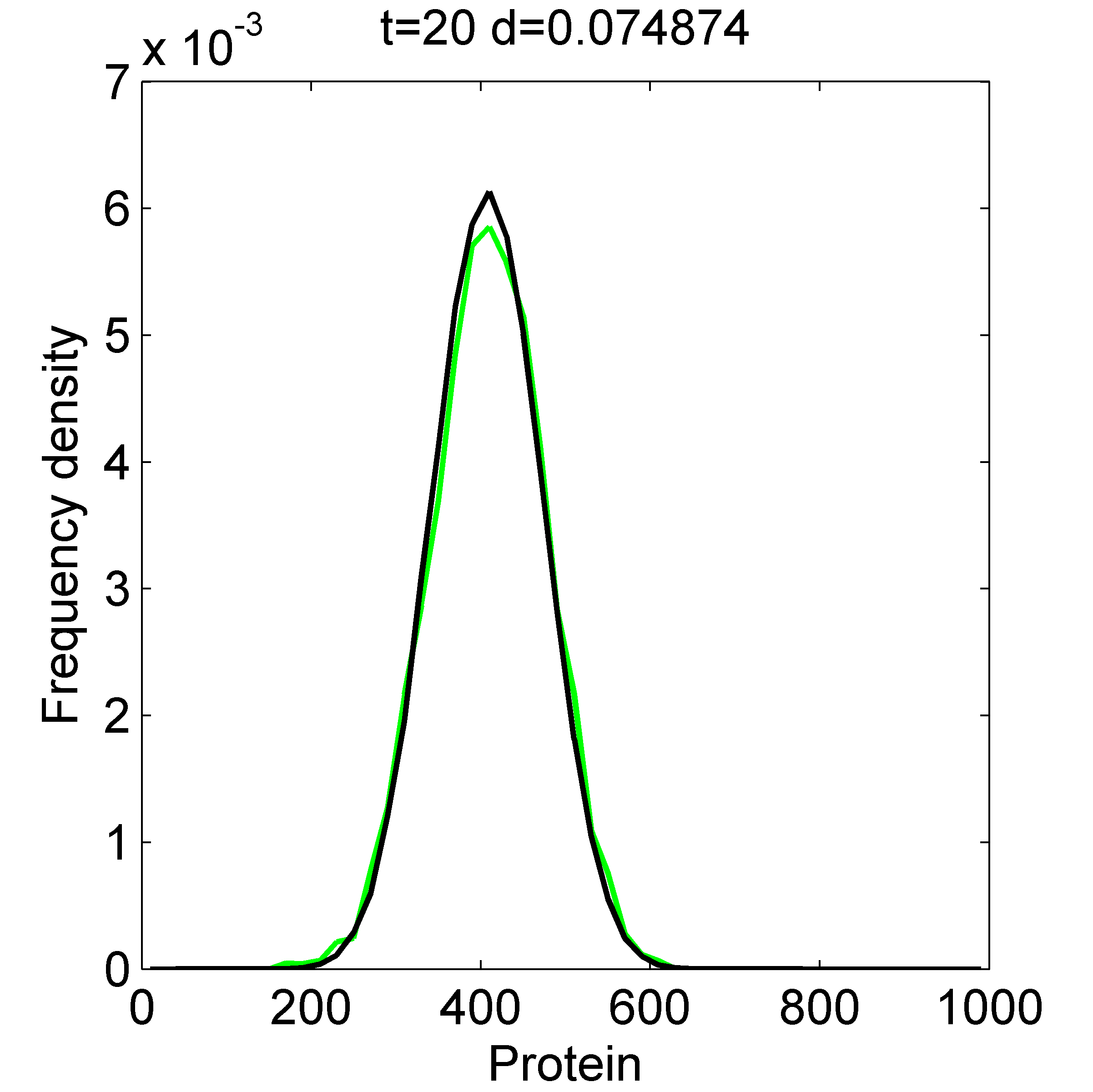}

\caption{ \label{fig1}
 Histograms of protein  for the second gene,
 produced by the Monte-Carlo method (green lines) and
 by the Push-forward method (black lines) for the model $M_1$.
 The comparison is quantified by the distance $d$ defined by \eqref{distance}.
  }
\end{figure}

\begin{figure}[h!]

\includegraphics[width=0.5\textwidth]{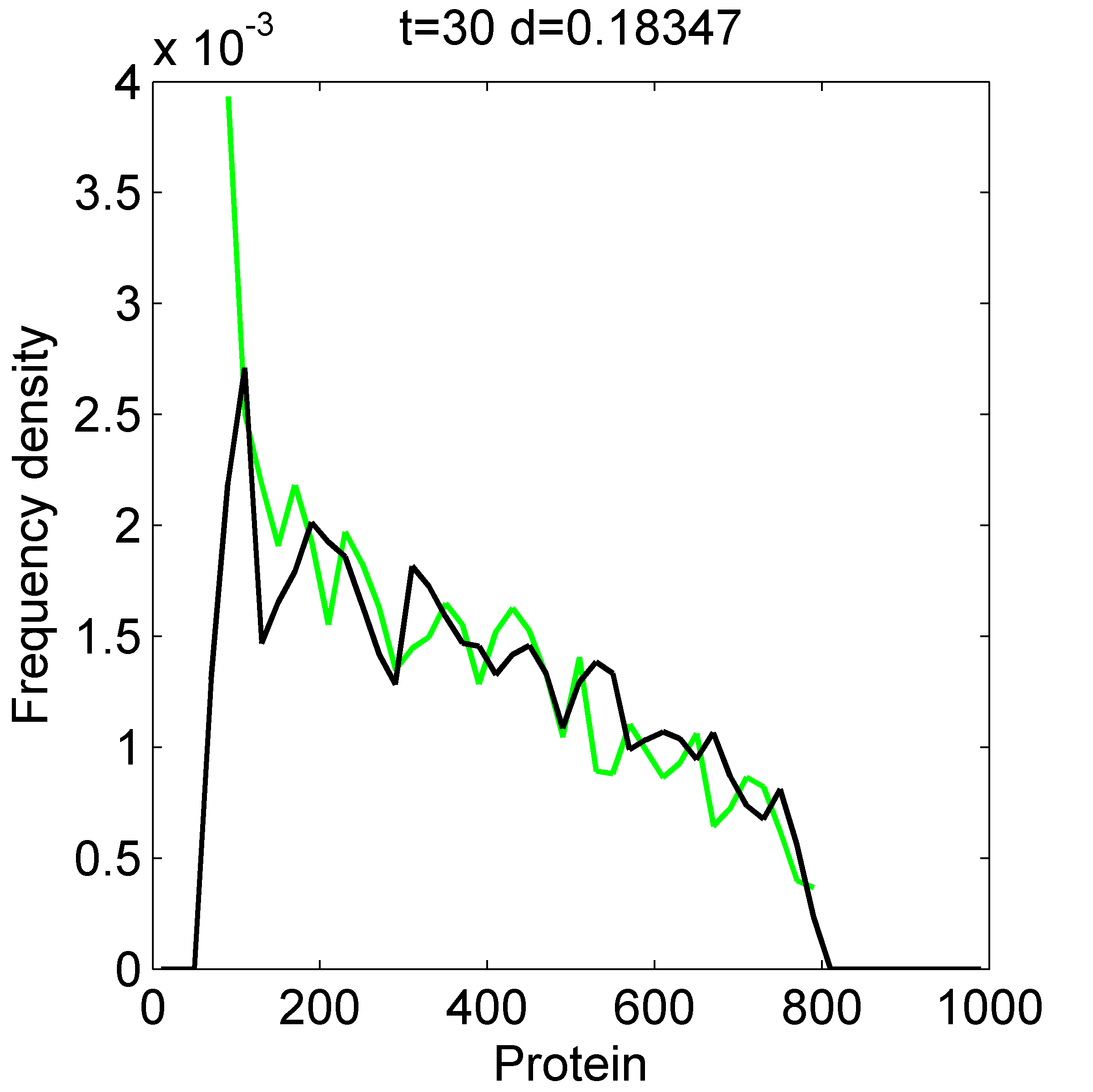}
\includegraphics[width=0.5\textwidth]{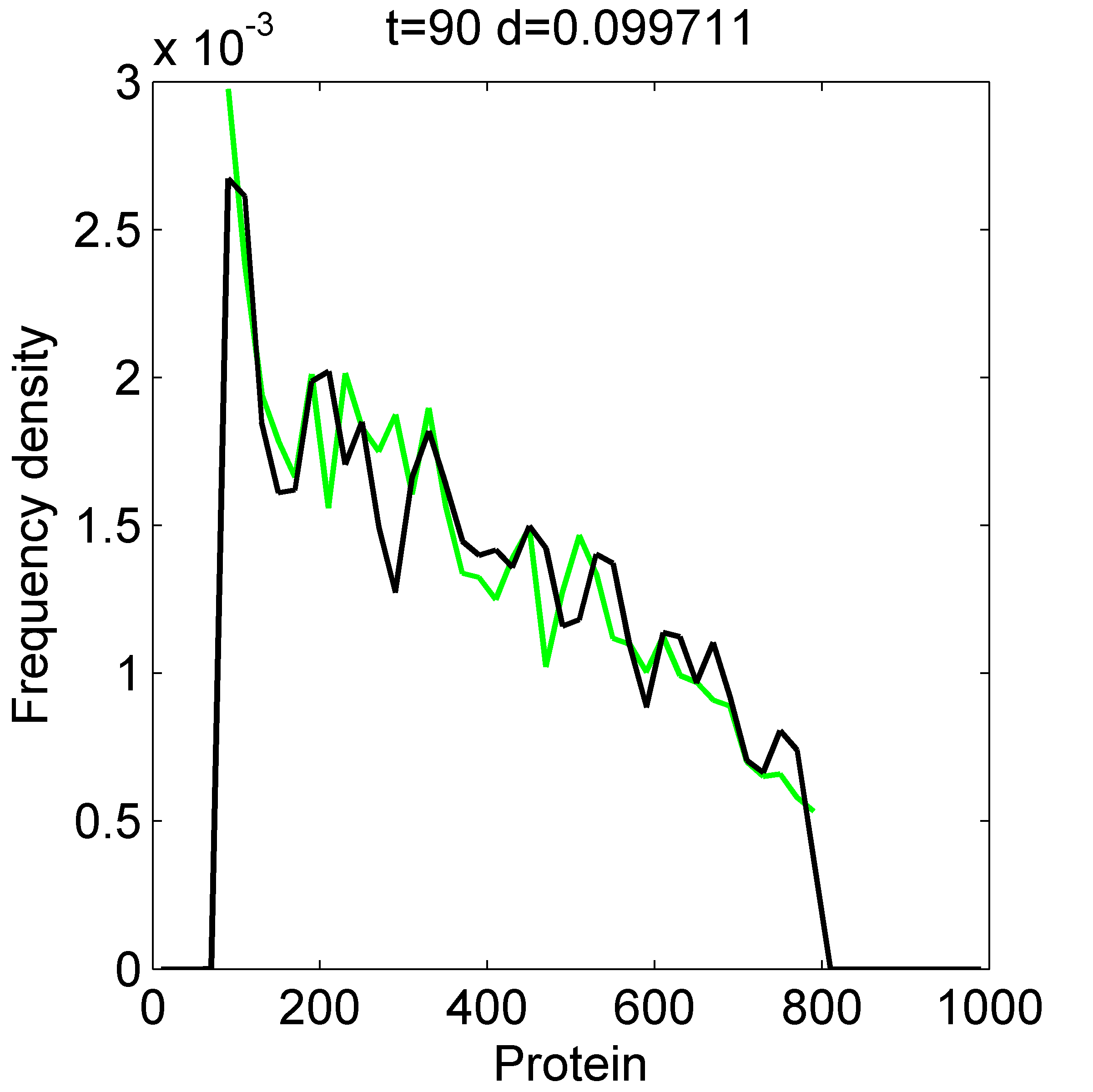}
\includegraphics[width=0.5\textwidth]{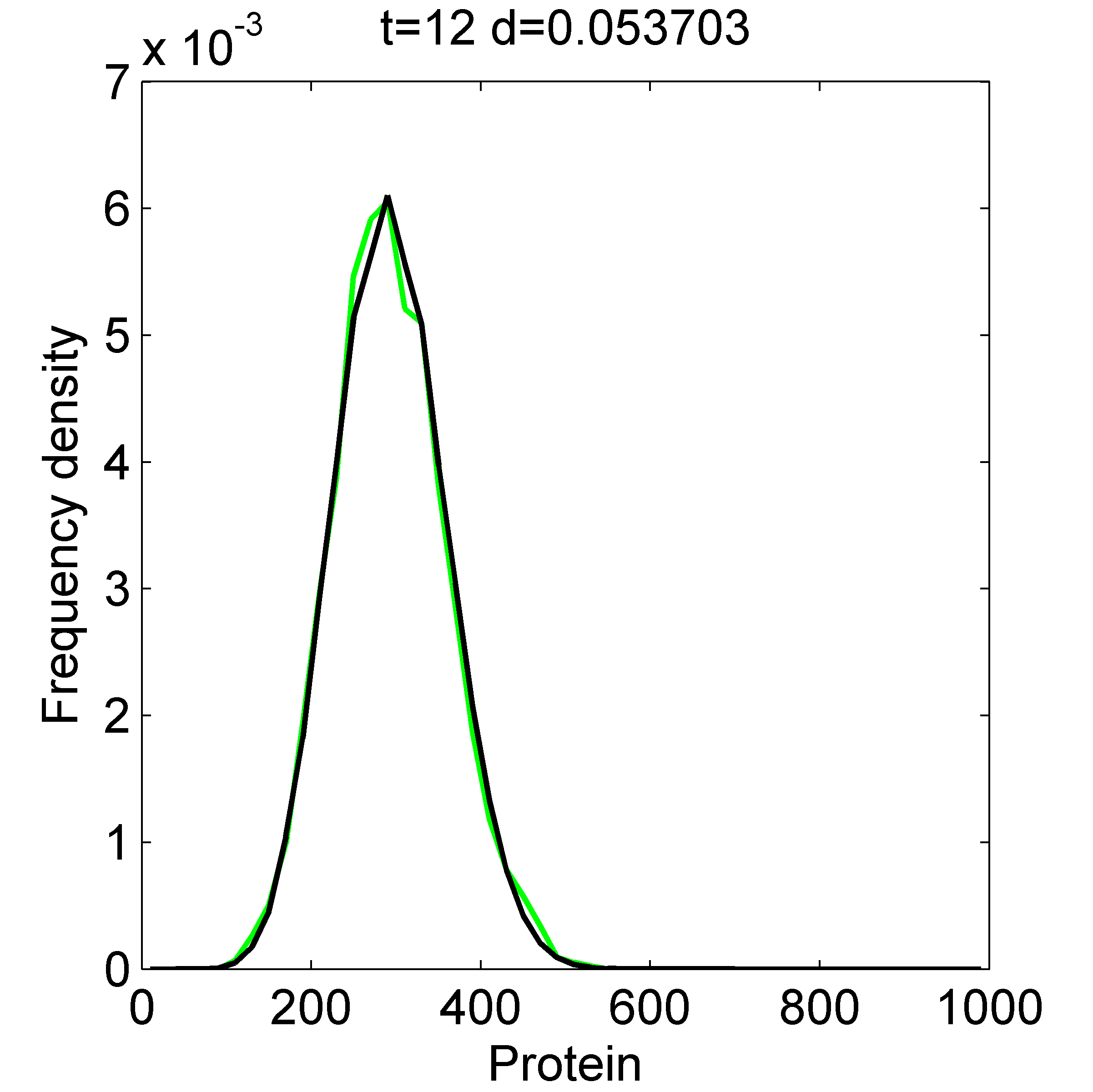}
\includegraphics[width=0.5\textwidth]{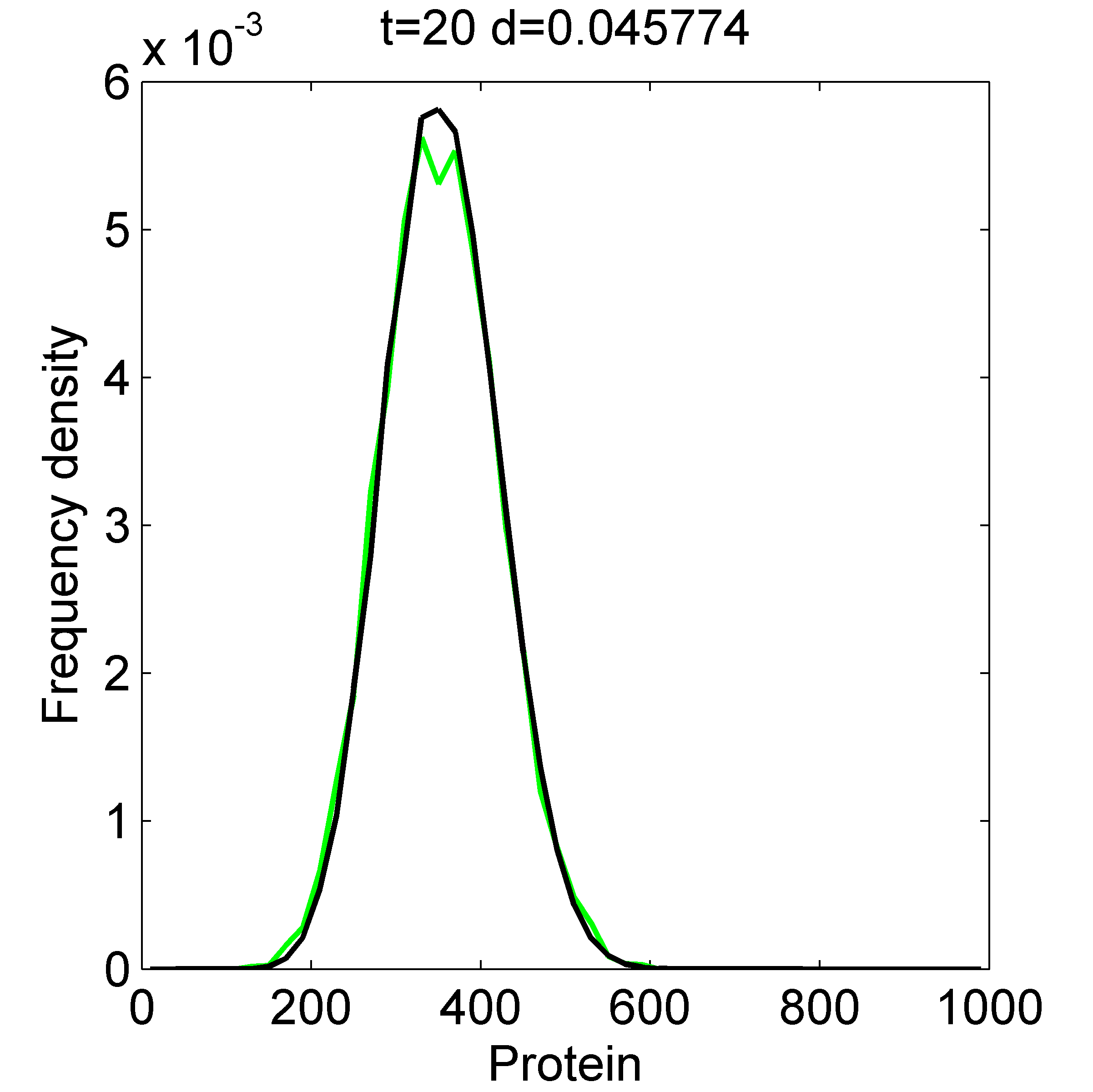}

\caption{ \label{fig2}
 Histograms of protein  for the second gene,
 produced by the Monte-Carlo method (green lines) and
 by the Push-forward method (black lines) for the model $M_6$.
 The comparison is quantified by the distance $d$ defined by \eqref{distance}.
  }
\end{figure}

The execution times are provided in the Table~\ref{table1}.
The comparison of the probability distributions
are illustrated in the Figures~\ref{fig1},\ref{fig2}.
 In order to quantify the relative difference between methods we use the $L^1$ distance
between distributions. More precisely, if $p(x)$ and $\tilde p(x)$ are probability density functions
to be compared, the distance between distributions is
\begin{equation}
 d = \int | p(x) - \tilde p(x) | \,dx.
 \label{distance}
 \end{equation}

\begin{table}
\begin{center}
\begin{tabular}{|c|c|c|}
\hline
Model  &  Monte-Carlo [min]  & Push-forward [s]  \\
\hline
$M1$ slow-slow & 45     &  20 \\
$M1$  fast-fast & 74    &  30 \\
$M2$  slow-slow & 447   &  20 \\
$M2$  fast-fast & 758   &  30 \\
\hline
\end{tabular}
\end{center}
\caption{\label{table1} Execution times for different methods.
All the methods were implemented in
Matlab R2013b running on a single core (multi-threading inactivated) of a Intel i5-700u 2.5 GHz processor.
The Monte-Carlo method computed the next jump waiting time using the analytical solution of Eq.\ref{T1} for $M_1$
and the numerical solution of Eq.\ref{T1} for $M_2$. The push-forward method used analytic solutions for
mRNA and protein trajectories from \eqref{eqyt},\eqref{eqxt}, and numerical computation of the integrals
in Eq.\ref{pi2}, for both models.
}
\end{table}

\section{Discussion and conclusion}
Combining direct simulation of PDMP gene network models
and analytic formulas for the ODE flow represents an effective, easy to implement
method for computing time dependent MPD of these models. However, the precision
of the Monte-Carlo estimates of the distributions increases like
$\sqrt{MC}$, where $MC$ is the number of Monte-Carlo samples. For this reason,
the execution time of this method, although smaller compared to PDMP simulation
methods that implement numerical resolution of the ODEs such as reported in
\cite{lin2018efficient} (data not shown), is large compared to
deterministic methods such as the push-forward method.

The push-forward method represents an effective alternative to Monte-Carlo methods,
ensuring reduced execution time. With respect to an earlier implementation of this method
in \cite{innocentini2016protein} we used promoter states instead of mRNA copy numbers
as discrete variables of the PDMP. As a consequence, the number of discrete states is
lower and we can afford increasing the number $M$ of time subdivisions. Compared to
the similar work in \cite{innocentini2018time} we used second moments of the protein
distribution which took into account the correlation of the promoter states and
lead to increased accuracy in the case of nonlinear regulation.
We proved rigorously the convergence of the distributions calculated with the
push-forward method to the exact distributions of the PDMP.
However, the push-forward method is an approximate method, and its accuracy relies on the
careful choice of the time and space steps, namely of the integers
$M$, $n_t$, $n_x$, $n_y$. We will present elsewhere error estimates allowing an optimal choice of
these parameters.
Although the protein moments and the exponential transition rate matrix $\vec{\Pi}$ can be computed
numerically, the effectiveness of the push-forward method is increased when analytic expressions are
available for these quantities. In this paper, these expressions were computed for
particular cases. In the future, we will provide expressions, as well
as symbolic computation tools to compute these quantities in more general cases.
We situate our findings in the broader effort of the community
to produce new effective tools for computational biology
by combining numerical and symbolic methods.

\section*{Appendix1: mean and variance of the protein}

We compute here the mean and the variance of the protein synthesized by a constitutive
promoter (gene 1 of models $M_1$ and $M_2$).

We start with
\begin{equation}
x(t) = x(0) e^{-at} + b \int_0^t \left[y(0) e^{-\rho t'} + \int_0^{t'} \frac{k_0 + (k_1-k_0)s(t'')}{\rho}e^{\rho(t''-t')}dt''\right]e^{a(t'-t)}dt', \label{a11}
\end{equation}
where $s(t)=0$ if the promoter is OFF and $s(t)=1$ if the promoter is ON at the time
$t$.

Eq.\ref{a11} leads to
\begin{equation}\begin{split}
x(t) &= x(0) e^{-at} + b y(0) \frac{e^{-\rho t}-e^{-a t}}{a-\rho}+
\frac{bk_0}{\rho}\left( \frac{1-e^{-at}}{a}-\frac{e^{-\rho t}-e^{-at}}{a-\rho}\right)
+ \\
&+\frac{b(k_1-k_0)}{\rho} \int_0^t \left[\int_0^{t'} s(t'') e^{\rho t''}dt'' \right]e^{(a-\rho)t'}e^{-at}dt'.
\end{split}\label{a12}
\end{equation}
From \eqref{a12} it follows
\begin{equation}\begin{split}
\esp{x(t)} &= \esp{x(0)} e^{-at} + b \esp{y(0)} \frac{e^{-\rho t}-e^{-a t}}{a-\rho}+
\frac{bk_0}{\rho}\left( \frac{1-e^{-at}}{a}-\frac{e^{-\rho t}-e^{-at}}{a-\rho}\right)
+ \\
&+\frac{b(k_1-k_0)}{\rho} \int_0^t \left[\int_0^{t'} \esp{s(t'')} e^{\rho t''}dt'' \right]e^{(a-\rho)t'}e^{-at}dt'.
\end{split}\label{a13}
\end{equation}
The promoter state variable $s(t)$ follows the master equation
\begin{equation}
\D{\proba{s(t)=1}}{t} = f(1-\proba{s(t)=1}) - (f+h)\proba{s(t)=1},
\label{a14}
\end{equation}
that has the solution
\begin{equation}
\esp{s(t)} = \proba{s(t)=1}= (p10-p_1)e^{-\rho \epsilon t} + p_1,
\label{a15}
\end{equation}
where $p10 = \proba{s(0)=1}$, $\epsilon = (f+h)/\rho$, and $p_1 = f/(f+h)$.
Using straightforward algebra, we find
\begin{equation}
\esp{x(t)} = M_0+M_1 e^{-at}+ M_2e^{-\rho t}+ M_3e^{-\epsilon t},
\label{a16}
\end{equation}
where
\begin{eqnarray}
M_0 &=& b/a (k_0 + (k_1-k_0) p_1), \\
M_1 &=& \esp{x(0)} - \frac{b\esp{y(0)}}{a-\rho}+ \frac{bk_0}{a(a-\rho)}
+ \frac{b(k_1-k_0)(p_{10}-p_1)}{(a-\rho\epsilon)(a-\rho)}+\frac{b(k_1-k_0)p_1}{a(a-\rho)}, \\
M_2 &=& \frac{b\esp{y(0)}}{a-\rho} - \frac{bk_0}{a-\rho} - \frac{b(k_1-k_0)(p_{10}-p_1)}{\rho(1-\epsilon)(a-\rho)} - \frac{b(k_1-k_0)p_1}{a-\rho}, \\
M_3 &=& \frac{b(k_1-k_0)(p_{10}-p_1)}{\rho(1-\epsilon)(a-\epsilon \rho)}.
\label{a17}
\end{eqnarray}
From \eqref{a12} we find also
\begin{equation}\begin{split}
&Var(x(t)) = Var(x(0)) e^{-2at} + b^2 Var(y(0)) \left( \frac{e^{-\rho t}-e^{-a t}}{a-\rho} \right)^2 +\left( \frac{b(k_1-k_0)}{\rho} \right)^2
e^{-2at} \times \\
&\times  \int_0^t \int_0^t dt_2dt_4\left[ \int_0^{t_2}\int_0^{t_4} (\esp{s(t_1)s(t_3)}-\esp{s(t_1)}\esp{s(t_3)})
e^{\rho t_1}e^{\rho t_3}dt_1dt_3 \right]e^{(a-\rho)t_2}e^{(a-\rho)t_4}.
\end{split}\label{a18}
\end{equation}
We have considered here that $x(0)$, $y(0)$ are uncorrelated, but
more general expressions can be obtained.

In order to compute the two times covariance $\esp{s(t_1)s(t_3)}-\esp{s(t_1)}\esp{s(t_3)}$ we combine the tower property of the
conditional expectation with the Markov property satisfied by $s(t)$. More precisely,
for $t_1 \geq  t_3$ we find
$\esp{s(t_1)s(t_3)}  = \esp{\esp{s(t_1)s(t_3)|s(t_3}} =
\esp{((s(t_3)-p_1)e^{-\rho\epsilon(t_1-t_3)}+p_1 )s(t_3)}$ and
$\esp{s(t_1)}\esp{s(t_3)}  =
\esp{\esp{s(t_1)|s(t_3}} \esp{s(t_3)}=
((\esp{s(t_3)}-p_1)e^{-\rho\epsilon(t_1-t_3)}+p_1 )\esp{s(t_3)}$. Then, it follows
\begin{equation}
\esp{s(t_1)s(t_3)} - \esp{s(t_1)}\esp{s(t_3)}  =
Var[s(t_3)] e^{-\rho\epsilon(t_1-t_3)}.
\label{a19}
\end{equation}
$s(t_3)$ is a Bernoulli variable, therefore $Var[s(t_3)] = \esp{s(t_3)} (1 - \esp{s(t_3)})$.
From \eqref{a19} and \eqref{a15} it follows
\begin{equation}
\begin{split}
&\esp{s(t_1)s(t_3)}-\esp{s(t_1)}\esp{s(t_3)} = p_1(1-p_1) e^{-\rho\epsilon(t_1-t_3)} +
(1-2p_1)(p10-p_1)e^{-\rho\epsilon t_1 }- \\ & - (p10-p_1)^2 e^{-\rho\epsilon(t_1+ t_3)},\text{ for }
t_1 \geq t_3.
\end{split}
\label{a20}
\end{equation}
Similarly, one gets
\begin{equation}
\begin{split}
&\esp{s(t_1)s(t_3)}-\esp{s(t_1)}\esp{s(t_3)} = p_1(1-p_1) e^{-\rho\epsilon(t_3-t_1)} +
(1-2p_1)(p10-p_1)e^{-\rho\epsilon t_3 } -\\ &- (p10-p_1)^2 e^{-\rho\epsilon (t_1+t_3)}, \text{ for }
t_3 \geq t_1.
\end{split}\label{a20p}
\end{equation}
The domain of the multiple integral in \eqref{a18} should be split in two sub-domains
corresponding to $t_2 < t_4$ and to $t_2 > t_4$. Each of these sub-domains
should be subdivided into two smaller sub-domains corresponding to $t_3 > t_1$ and $t_1 < t_3$. Symmetry arguments imply  that the
integrals on $t_2 < t_4$ and on $t_4 < t_2$ are equal, which allows us to perform
the calculation of the integral on only two sub-domains, instead of four.
After some algebra we find
\begin{eqnarray}
&Var(x(t)) = Var(x(0)) e^{-2at} + b^2 Var(y(0)) \left( \frac{e^{-\rho t}-e^{-a t}}{a-\rho} \right)^2
- \left[\frac{(p_{10}-p_1)(k_1-k_0)b}{\rho(1-\epsilon)}\left( \frac{e^{-\rho\epsilon t}-e^{-a t}}{a-\rho\epsilon} - \frac{e^{-\epsilon t}-e^{-a t}}{a-\rho\epsilon}\right) \right]^2
 \notag \\
+&\frac{p_1(1-p_1)(k_1-k_0)^2 b^2}{\rho^2} (V_0
+V_1 e^{-(a+\rho\epsilon)t} + V_2e^{-\rho(1+\epsilon)t}+
V_3 e^{-2at} + V_4 e^{-(a+\rho)t}+ V_5 e^{-2\rho t}) +\notag \\
+&\frac{(1-2p_1)(p10-p_1)(k_1-k_0)^2 b^2}{\rho^2}(
V_6 e^{-\rho \epsilon t} +V_7 e^{-(a-\rho \epsilon) t}
+V_8 e^{-\rho (\epsilon+1) t}
+V_9 e^{-(a+\rho \epsilon) t}
+V_{10} e^{-\rho t} +
V_{11} e^{-2\rho t}),
\end{eqnarray}
where
\begin{eqnarray}
V_0 &= \frac{a+(\epsilon+1)\rho}{a(a+\rho\epsilon)(a+\rho)(\epsilon+1)},
V_1 &= -\frac{2}{(a^2-\rho^2\epsilon^2)(a-\rho)(\epsilon-1)},\notag \\
V_2 &= \frac{2}{(a-\rho\epsilon)(a-\rho)(\epsilon^2-1)},
V_3 &= \frac{1}{a(a-\rho\epsilon)(a-\rho)^2}, \notag \\ V_4&=\frac{2(a+(1-2\epsilon)\rho)}{(a-\rho\epsilon)(a-\rho)^2(a+\rho)(\epsilon-1)},
V_5 &= -\frac{1}{(\epsilon-1)(a-\rho)^2}, \notag \\
V_6 &= -\frac{2(2a+(2-\epsilon)\rho)}{a(\epsilon-2)(2a-\rho\epsilon)(a+(1-\epsilon)\rho)},
V_7 &= \frac{2}{(1-\epsilon)a(a-\rho)(a-\rho\epsilon)}, \notag \\
V_8 &= \frac{2}{(a-\rho\epsilon)(a-\rho)(\epsilon-1)},
V_9 &= \frac{2(a+(1-2\epsilon)\rho)}{(a-\rho)^2(\epsilon-1)(a-\rho\epsilon)(a+(1-\epsilon))}, \notag \\
V_{10} &= \frac{2}{(a-\rho)^2(2-\epsilon)(1-\epsilon)},
V_{11} &= \frac{2}{(a-\rho)^2(2a-\rho\epsilon)(a-\rho\epsilon)}.
\end{eqnarray}

\section*{Appendix2: details of the derivation of \eqref{eqyt},\eqref{eqxt}}
$x(t)$ and $y(t)$ satisfy the following system of equations
\begin{eqnarray}
\D{x}{t} &=& b y - a x \notag \\
\D{y}{t} &=& k_0 + (k_1-k_0) s - \rho y
\label{a21}
\end{eqnarray}
For simplification, we rescale variables and parameters
$t \to t \rho$, $k_i \to k_i / \rho$, $a \to a/\rho$, $b \to b/\rho$
and obtain
\begin{eqnarray}
\D{x}{t} &=& b y - a x \notag \\
\D{y}{t} &=& k_0 + (k_1-k_0) s - y
\label{a22}
\end{eqnarray}
From \eqref{a22} it follows
$y(\tau) = y(0) e^{-\tau} + \int_0^{\tau} d\tau' e^{-(\tau-\tau')}[k_0 + (k_1-k_0)s(\tau')] = y(0) e^{-\tau} + \sum_{j=1}{M-1}\int_{\tau_j}^{\tau_{j+1}}d\tau'
e^{-(\tau-\tau')}[k_0 + (k_1-k_0)s(\tau_j)]=
y(0) e^{-\tau}  + k_0(1-e^{-\tau}) + (k_1-k_0)\sum_{j=0}{M-1}e^{-\tau}(e^{\tau_{j+1}}-e^{\tau_j})s(\tau_j)$ and hence
\eqref{eqyt}.

\begin{figure}[h!]
	\centering
			\includegraphics[width=50mm]{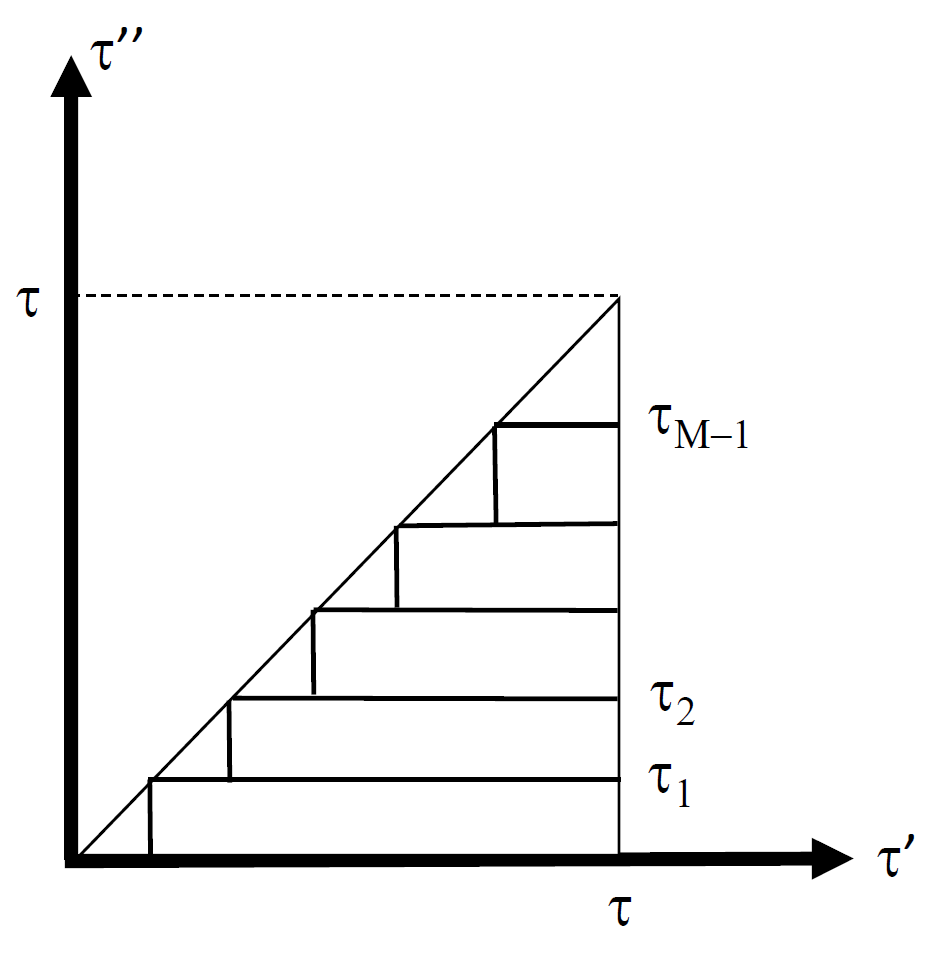}
		\caption{Decomposition of the integration domain for computing the integral $I$.\label{fig:domain}}
	\end{figure}

From \eqref{a22} we also obtain
$x(\tau) = x(0) e^{-a\tau} + b\int_0^{\tau} e^{a(\tau'-\tau)} y(\tau') d \tau'=
x(0) e^{-a\tau} + \frac{by(0)}{a-1}(e^{-\tau}-e^{-a\tau})
+ b\int_0^\tau d\tau' e^{a(\tau'-\tau)} \int_0^{\tau'} d\tau'' e^{-(\tau'-\tau'')}
(k_0 + (k_1-k_0) s(\tau''))=
x(0) e^{-a\tau} + \frac{by(0)}{a-1}(e^{-\tau}-e^{-a\tau})+
bk_0 \left( \frac{1-e^{-a\tau}}{a} + \frac{e^{-a(\tau-\tau_0)} - e^{-(\tau-\tau_0)} }{a-1}   \right)
+ b(k_1-k_0) I,
$
where
$$I = \int_0^{\tau} d\tau' e^{a(\tau'-\tau)} \int_0^{\tau'} d\tau'' e^{-(\tau'-\tau'')}s(\tau'').$$
In order to compute the integral $I$ we decompose the triangular integration domain
into $M-1$ rectangles and $M$ triangles on each of which  $s(\tau'')$ is constant,
as in Figure\ref{fig:domain}.

The contribution of each rectangle to the integral $I$ is
$\int_{\tau_i}^{\tau} d\tau' e^{a(\tau'-\tau)} \int_{\tau_{i-1}}^{\tau_i} d\tau'' e^{-(\tau'-\tau'')}s(\tau_{i-1})=e^{-a\tau} \frac{e^{(a-1)\tau}-e^{(a-1)\tau_i}}{a-1}
(e^{\tau_i}-e^{\tau_{i-1}})s(\tau_{i-1})$.

The contribution of each triangle to the integral $I$ is
$\int_{\tau_{i-1}}^{\tau_i} d\tau' e^{a(\tau'-\tau)} \int_{\tau_{i-1}}^{\tau'} d\tau'' e^{-(\tau'-\tau'')}s(\tau_{i-1})=
e^{-a\tau}s(\tau_{i-1}) \left(
\frac{e^{a\tau_i}-e^{a\tau_{i-1}}}{a}
-e^{\tau_{i-1}}
\frac{e^{(a-1)\tau_i}-e^{(a-1)\tau_{i-1}}}{a}
\right)$.

It follows that
$I = \sum_{i=1}^{M-1} e^{-a\tau} \frac{e^{(a-1)\tau}-e^{(a-1)\tau_i}}{a-1}
(e^{\tau_i}-e^{\tau_{i-1}})s(\tau_{i-1})+
\sum_{i=1}^M e^{-a\tau} s(\tau_{i-1}) \left(
\frac{e^{a\tau_i}-e^{a\tau_{i-1}}}{a}
-e^{\tau_{i-1}}
\frac{e^{(a-1)\tau_i}-e^{(a-1)\tau_{i-1}}}{a}
\right).$

Noting that the first sum in the expression of $I$ can go to $i=M$ (the M-th
term is zero) we obtain \eqref{eqxt}.

\section*{Appendix3: proof of the Theorem~\ref{theorem1}}

The proof the Theorem 1 relies on the following Lemma:
\begin{lemma}\label{lemma1}
Let $(x(t),y(t),s(t))$ be a realization of the PDMP such that $s(\tau_j) = s_j$
for all $j \in [0,M-1]$ and let $(x_M(t),y_M(t))$ be the push-forward solutions
computed with \eqref{eqyt},\eqref{eqxt} considering that $s(t)$ is piecewise
constant on the intervals $[\tau_{i-1},\tau_{i}]$. Then, under the conditions
of Theorem 1, $\proba{|y(t)-y_M(t)|> \epsilon} \to 0$ and $\proba{|x(t)-x_M(t)|> \epsilon} \to 0$ when $M \to \infty$, for any
$\epsilon >0$ and for all $t\in[0,\tau]$.
\end{lemma}

We prove this Lemma for $y(t)$. The proof for $x(t)$ follows the same principles.

According to the constructive definitions of PDMP (see Section 2.1), and
considering that $\lambda(\vec{x}(t),\vec{y}(t),s(t)) < A$, for all $t \in [0,\tau]$ (this follows from the continuity of $\lambda$ and the
boundedness of $\vec{x}(t)$, and $\vec{y}(t)$) then, with probability one,
$s(t)$ has a finite number of jumps inside the interval $[0,\tau]$. Consider
that for the  $i^\text{th}$ gene  there are $k$ jumps such that $s_i$ changes from
ON to OFF or from OFF to ON. The positions of these jumps are
$\tau^*_{j_l} \in [\tau_{j_l},\tau_{j_l+1})$, for $l\in[1,k]$. Using \eqref{eqyt}
it follows
$|y_i(t)-y_{i,M}(t)| \leq e^{-\rho \tau}\frac{k_1-k_0}{\rho}
\sum_{l=1}{k}|e^{\rho \tau^*_{j_l}}- e^{\rho \tau_{j_l}}| \leq \frac{k_1-k_0}{\rho}
\frac{k_1-k_0}{\rho} \sum_{l=1}{k} (\tau_{j_{l+1}}-\tau_{j_l}) < \frac{k_1-k_0}{\rho} \frac{k C}{M}$.

For constitutive promoters the number of jumps $k$ of the promoter of the
gene $i$ inside $[0,\tau]$ has a mean $\esp{k} = \frac{\tau}{h_i+f_i}$. Slightly
more complex, but finite bounds, can be obtained for a regulated gene as well.

Using Markov's inequality we find that
$\proba{\frac{k_1-k_0}{\rho} \frac{k C}{M} > \epsilon} \leq
\frac{k_1-k_0}{\rho} \frac{C \tau}{\epsilon(h_i+f_i)M}$. It follows
that $\proba{|y(t)-y_M(t)|> \epsilon} \to 0$ when $M \to \infty$, for any
$\epsilon >0$.

The proof of the Theorem 1 follows from the Lemma~\ref{lemma1} because
by construction, the promoter states have the same distribution in the push-forward and PDMP schemes, and the convergence in probability of the
mRNAs and of the proteins implies the convergence in distribution of these variables.


\end{document}